\documentstyle[twocolumn,aps,floats]{revtex}

\begin{document}

\draft \tolerance = 10000

\setcounter{topnumber}{1}
\renewcommand{\topfraction}{0.9}
\renewcommand{\textfraction}{0.1}
\renewcommand{\floatpagefraction}{0.9}
\newcommand{\br}{{\bf r}}

\twocolumn[\hsize\textwidth\columnwidth\hsize\csname
@twocolumnfalse\endcsname

\title{What Dimensions Do the Time and Space Have: Integer or Fractional?}
\author{L.Ya.Kobelev }
\address{Department of  Physics, Urals State University \\ Lenina
Ave., 51, Ekaterinburg 620083, Russia \\ Electronic address:
leonid.kobelev@usu.ru   }

\maketitle

\begin{abstract}
A theory of time and space with fractional dimensions (FD) of time
and space ($d_{\alpha},\; \alpha=t,{\bf r})$ defined on
multifractal sets is proposed. The FD is determined (using
principle of minimum the functionals of FD) by the energy
densities of Lagrangians of known physical fields. To describe
behaviour of functions defined on multifractal sets the
generalizations of the fractional Riemann-Liouville derivatives
$D_{t}^{d(t)}$ are introduced with the order of differentiation
(depending on time and coordinate) being equal the value of
fractional dimension. For $d_{t}=const$ the generalized fractional
derivatives (GFD) reduce to ordinary Riemann-Liouville integral
functionals, and when $d_{t}$ is close to integer, GFD can be
represented by means of derivatives of integer order. For time and
space with fractional dimensions a method to investigate the
generalized  equations of theoretical physics by means of GFD is
proposed. The Euler equations defined on multifractal sets of time
and space are obtained using the principle of the minimum of FD
functionals. As an example, a generalized Newton equation is
considered and it is shown that this equation coincide with the
equation of classical limit of general theory of relativity for
$d_{t} \to 1$. Several remarks concerning existence of repulsive
gravitation are discussed. The possibility of geometrization all
the known physical fields and forces in the frames of the fractal
theory of time and space is demonstrated.
\end{abstract}

 \vspace{1cm}

]

\section{Introduction}
The problem concerning the nature of space and time is one of the
most interesting problems of the modern physics. Are the space and
time continuous? Why is time irreversible? What dimensions do
space and time have? How is the nature of time in the equations of
modern physics is reflected? Different approaches (quantum
gravity, irreversible thermodynamics, synergetics and others)
provide us with different answers to these questions. In this
paper the hypothesis about a nature of time and space based on an
ideas of the fractal geometry \cite{man} is offered. The
corresponding mathematical methods this hypothesis makes use of
are based on using the idea about fractional dimensions (FD) as
the main characteristics of time and space and in connection with
this the generalization of the Riemann-Liouville fractional
derivatives are introduced. The method and theory are developed to
describe dynamics of functions defined on multifractal sets of
time and space with FD.

Following \cite{kob1}, we will consider both time and space as an
only material fields existing in the Universe and generating all
other physical fields. Assume that each of them consists of a
continuous, but not differentiable bounded set of small elements.
Let us suppose a continuity, but not a differentiability, of sets
of small time intervals (from which time consist) and small space
intervals (from which space consist). First, let us consider set
of small time intervals $S_{t}$ (for the set of small space
intervals the way of reasoning is similar). Let time be defined on
multifractal set of such intervals (determined on the carrier of a
measure ${\mathcal R}_{t}^{n}$ ). Each of intervals of this set
(further we use the approximation in which the description of each
multifractal interval of these sets will be characterized by
middle time moment $t$ and refer to each of these intervals as
"points") is characterized by global fractal dimension (FD)
$d_{t}({\mathbf r}(t),t)$, and for different intervals FD are
different( because of the time dependence and spatial coordinates
dependence of $d_{t}$ ). For multifractal sets $S_{t}$ (or
$S_{r}$) each set is characterized by global FD of this set and by
local FD of this set( the characteristics of local FD of time and
space sets in this paper we do not research). In this case the
classical mathematical calculus or fractional (say, Riemann -
Liouville) calculus \cite{sam} can not be applied to describe a
small changes of a continuous function of physical values $f(t)$,
defined on time subsets $S_{t}$, because the fractional exponent
depends on the coordinates and time. Therefore, we have to
introduce integral functionals (both left-sided and right-sided)
which are suitable to describe the dynamics of functions defined
on multifractal sets (see [1]). Actually, this functionals are
simple and natural generalization the Riemann-Liouville fractional
derivatives and integrals:
\begin{equation} \label{1}
D_{a+,t}^{d}f(t)=\left( \frac{d}{dt}\right)^{n}\int_{a}^{t}
\frac{f(t^{\prime})dt^{\prime}}{\Gamma
(n-d(t^{\prime}))(t-t^{\prime})^{d(t^{\prime})-n+1}}
\end{equation}
\begin{equation} \label{2}
D_{b-,t}^{d}f(t)=(-1)^{n}\left( \frac{d}{dt}\right)
^{n}\int_{t}^{b}\frac{f(t^{\prime})dt^{\prime}}{\Gamma
(n-d(t^{\prime}))(t^{\prime}-t)^{d(t^{\prime})-n+1}}
\end{equation}
where $\Gamma(x)$ is Euler's gamma function, and $a$ and $b$ are some
constants from $[0,\infty) $. In these definitions, as usually,
$n=\{d\}+1$ , where $\{d\}$ is the integer part of $d$ if $d\geq 0$ (i.e.
$n-1\le d<n$) and $n=0$ for $d<0$. Functions under the integral sign we
will consider to be generalized functions defined on the space of finite
functions \cite{gel}. Similar expressions can be written down for GFD of
functions $f({\bf r},t)$ with respect to spatial variables ${\mathbf r}$,
with $f({\bf r},t)$ being defined on the elements of set $S_{\bf r}$ whose
dimension is $d_{\bf r}$.

For an arbitrary $f(t)$ it is useful to expand the generalized function
$1/(t-t')^{\varepsilon(t')}$ under the integral sign in
(\ref{1})-(\ref{2}) into a power series in $\varepsilon(t')$ when
$d=n+\varepsilon,\, \varepsilon \to +0$ and write
\begin{eqnarray}      \label{1a}
D_{a+,t}^{d}f(t)=\left( \frac{d}{dt} \right)^{n} \int_{a}^{t}
\frac{f(t^{\prime})}{\Gamma(n-d(t^{\prime}))(t-t^{\prime}) } \\
\nonumber
\times\left(1+\varepsilon(t^{\prime})\ln(t-t^{\prime})+...\right)
dt^{\prime}
\end{eqnarray}
\begin{eqnarray}      \label{2a}
D_{b-,t}^{d}f(t)=(-1)^{n}\left( \frac{d}{dt}\right) ^{n}\int_{t}^{b}
\frac{f(t^{\prime})}{\Gamma(n-d(t^{\prime}))(t'-t) } \\ \nonumber
\times\left(1+\varepsilon(t^{\prime})\ln(t'-t)+...\right) dt^{\prime}
\end{eqnarray}
Taking into account that all functions here are real functions and $1/t =
P(1/t)\pm \pi i \delta(t)$, singular integrals here can be defined through
the rule
\begin{equation}     \label{3}
\int_{0}^{t}\frac{f(t')}{t-t'}dt'=af(t)
\end{equation}
where $a$ is a real regularization factor. A good agreement of
(\ref{1a})-(\ref{2a}) with the exact values given by expressions
(\ref{1})-(\ref{2}) can be obtained at large time by fitting the value of
$a$.

Instead of usual integrals and usual partial derivatives, in the frames of
multifractional time hypothesis it is necessary to use GFD operators to
describe small alteration of physical variables. These functionals reduce
to ordinary integrals and derivatives if space and time dimensions are
taken to be integer, and coincide with the Riemann-Liouville fractional
operators if $d_{i}=const$. If fractional dimension can be represented as
$d_{i}=n+\varepsilon_{i}({\bf r}(t),t),\,|\varepsilon|\ll 1$, it is  also
possible to reduce GFD to ordinary derivatives of integer order. Here we
show this only for the case when $d=1-\varepsilon <1$
\begin{eqnarray} \label{4}
D_{0+}^{1+\varepsilon}f(t)&=& \frac{\partial}{\partial t}
\int_{0}^{t}
\frac{\varepsilon(\tau)f(\tau)d\tau}{\Gamma(1+\varepsilon(\tau))
(t-\tau)^{1-\varepsilon(\tau)}} \\ \nonumber &\approx&
\frac{\partial}{\partial t} \int_{0}^{t}
\frac{\varepsilon(\tau)f(\tau)d\tau}{(t-\tau)^{1-\varepsilon(\tau)}}
\end{eqnarray}
Though for $\varepsilon \ne 0$ the last integral is well defined and is
real-valued, expanding it in power series in $\varepsilon$ leads to
singular integrals like (\ref{3})
$$A=\int_{0}^{t}\frac{\varepsilon(t')f(t')}{t-t'}dt'$$ To regularize this
integral we will consider it to be defined on the space of finite main
functions $\varphi(t')/2\pi i$ and take the real part of the common
regularization procedure
\begin{equation}     \label{5}
A=a\varepsilon(t)f(t)
\end{equation}
Thus we obtain
\begin{equation}   \label{6}
D_{0+}^{1+\varepsilon}f(t)=\frac{\partial}{\partial t}f(t)+
\frac{\partial}{\partial t}\left[a\varepsilon({\bf
r}(t),t)f(t)\right]
\end{equation}
where $a$ is a regularization parameter. For the sake of
independence of GFD from this constant it is useful in the
following to choose $\beta_{i}$ (on which $\varepsilon$ depends
linearily) proportional to $a^{-1}$. It can be shown that for
large $t$ the exact expressions for the terms in
(\ref{1})-(\ref{2}) proportional to $\varepsilon$ are very close
to the approximate expression given by (\ref{6}) provided a
special choice for the parameter $\alpha$ is is made
($t=t_{0}+(t-t_{0}),\,t-t_{0}\ll t_{0},\,\alpha \sim \ln t \sim
\ln t_{0}$)

\section{Equations of physical theories in multifractal time and space}

Equations describing dynamics of physical fields, particles and so
on can be obtained from the principle of minimum of fractional
dimensions functionals. To do this, introduce functionals of
fractional dimensions of space and time $F_{\alpha}(...
|d_{\alpha}({\bf r})),\, \alpha =t,{\bf r}$. These functionals are
quite similar to the free energy functionals,but now it is
fractional dimension (FD) that plays the role of an order
parameter (see also \cite{kob1}). Assume further that FD
$d_{\alpha}$ is determined by the Lagrangian densities
$L_{\alpha,i}, (i=1,2,...,\,\alpha=t,{\bf r})$ of all the fields
$\psi_{\alpha, i}$, describing the particles and $\Phi_{\alpha,i}$
describing the interactions in the point ({\bf r})
\begin{equation}    \label{7}
d_{\alpha}=d_{\alpha}[L_{\alpha, i}({\bf r},t)]
\end{equation}
Equations that govern $d_{\alpha}$ behavior can be found by minimizing
this functional and lead to the Euler's equations written down in terms of
GFD defined in (\ref{1})-(\ref{2})
\begin{equation}      \label{8}
D_{+, L_{\alpha, i}(x)}^{d_{\alpha}}d_{\alpha}- D_{-, x}^{d_{\alpha}}
D_{+, L^{\prime}_{\alpha, i}(x)}^{d_{\alpha}}d_{\alpha}=0
\end{equation}
Substitution in this equation GFD for usual derivatives and specifying the
choice for $F$ dependence on $d_{\alpha}$ and relations between
$d_{\alpha}$ and $L_{\alpha}$ (the latter can correspond to the well known
quantum field theory Lagrangians) makes possible to write down the
functional dependence $F[L]$ in the form ($a,b,c$ are unknown functions of
$L$ or constants, $L_{0}$ is infinitely large density of the measure
carrier ${\mathcal R}^{n}$ energy)
\begin{eqnarray}   \label{9}  \nonumber
F(...|d_{\alpha})&=&\int dL_{\alpha}       \left\{
\frac{1}{2}[a(L_{\alpha})\frac{\partial d_{\alpha}}{\partial
L_{\alpha}}]^{2} \right. \\  &+& \left.
\frac{b(L_{\alpha})}{2}(L_{\alpha}-L_{\alpha, 0})d_{\alpha}^{2}+
c(L_{\alpha})d_{\alpha}  \right\}
\end{eqnarray}
or
\begin{eqnarray}  \label{10} \nonumber
F(...|d_{\alpha}) &=&  \int d^{4}L_{\alpha} \left\{
\frac{1}{2}[a(L_{\alpha})\frac{\partial d_{\alpha}}{\partial
L_{\alpha}}]^{2} \right. \\ &+& \left.
\frac{b(L_{\alpha})}{2}(L_{\alpha}-L_{\alpha, 0})d_{\alpha}^{2}+
\frac{1}{4}c(L_{\alpha})d^{4}_{\alpha} \right\}
\end{eqnarray}
The equations that determine the value of fractional dimension
follow from taking the variation of (\ref{9})-(\ref{10}) and read
\begin{equation}                    \label{11}
\frac{\partial}{\partial L}\left(a(L)\frac{\partial d_{t,\alpha}}{\partial
L}\right) +b(L)(L-L_{0})d_{\alpha}+c(L)d_{\alpha}^{2}=0
\end{equation}
or
\begin{equation}                    \label{12}
\frac{\partial}{\partial L_{\alpha}}\left(a(L_{\alpha})\frac{\partial
d_{t,\alpha}}{\partial L_{\alpha}}\right) +b(L_{\alpha})
(L_{\alpha}-L_{0,\alpha})d_{\alpha}^{2}+c(L_{\alpha})d_{\alpha}^{4}=0
\end{equation}
For nonstationary processes one have to substitute the time
derivative of $d_{\alpha}$ into the right-hand side of
Eqs.(\ref{11})-(\ref{12}). Neglecting the diffusion of
$d_{\alpha}$ processes in the space with energy densities  given
by the Lagrangians $L$ we can define
$L_{\alpha}-L_{\alpha,0}=\tilde{L}_{\alpha}\ll L_{\alpha, 0}$ with
$\tilde{L}_{\alpha}$ having sense of over vacuum energy density
and for the simplest case (\ref{11}) gives ($\alpha =t,\,
L_{t,i}\equiv L_{i}$)
\begin{equation}     \label{13}
d_{t}=\tilde{L}_{t}=1+\sum_{i}\beta_{i}L_{i}(t,{\mathbf
r},\Phi_{i},\psi_{i})
\end{equation}
More complicated dependencies of $d_{\alpha}$ on $L_{\alpha, i}$ are
considered in \cite{kob1}. Note that relation (\ref{13})  (and similar
expression for $d_{ {\mathbf r} }$ does not contain any limitations on the
value of $\beta_{i}L_{i}(t,{\mathbf r},\Phi_{i},\psi_{i})$ unless such
limitations are imposed on the corresponding Lagrangians, and therefore
$d_{t}$ can reach any whatever high or small value.

The principle of fractal dimension minimum, consisting in the requirement
for $F_{\alpha}$ variations to vanish under variation with respect to any
field, in this theory  produce the principle of energy minimum (for any
type of fractional dimension dependency on the Lagrangian densities). It
allows to receive Euler's-like equations with generalized fractional
derivatives for functions $f(y(x),y'(x))$, that describe behaviour of
physical value $f$ depending on physical variables $y$ and their
generalized fractional derivatives $y'=D^{d_{\alpha}}_{+,x}f$
\begin{equation}  \label{14}
\delta F_{t,y_{i}} \sim \delta d_{t, y_{i}} =0
\end{equation}
\begin{equation}\label{15}
\delta_{y_{i}}d_{\alpha}(f)=\delta_{y_{i}}L_{\alpha,i}(f)=0, \;\;\;\;\;\;
\alpha={\bf r},t
\end{equation}
\begin{equation}\label{16}
D^{d_{\alpha}}_{+,y_{i}(x)}f-
D^{d_{\alpha}}_{-,x}D^{d_{\alpha}}_{+,y_{i}'(x)}f=0
\end{equation}
The boundary conditions will have the form
\begin{equation}   \label{17}
D^{d_{\alpha}}_{+,y_{i}'(x)}f \Big|_{x_{0}}^{x_{1}}=0
\end{equation}
In these equations the variables $x$ stand for either $t$ or ${\bf
r}$ (the latter takes into account fractality of spatial
dimensions),
$y_{i}=\left\{\Phi_{i},\psi_{i}\right\},\,(i=1,2,...),\;
L_{\alpha, i}$ are the Lagrangian densities of the fields and
particles. Here $f$ can be of any mathematical nature (scalar,
vector, tensor, spinor, etc.), and modification of these equations
for functions $f$ of more complicated structure does not encounter
any principal difficulties. As Lagrangians $L_{\alpha, i}$ one can
choose any of the known in the theoretical physics Lagrangians of
fields and their sums, taking into account interactions between
different fields.

From Eq.(\ref{15}) it is possible to obtain generalizations of all known
equations of physics (Newton, Shroedinger, Dirac, Einstein equations and
etc.), and the similar equations for fractional space dimensions ($\alpha
= {\bf r}$). Such generalized equations extend the application of the
corresponding theories for the cases when time and space are defined on
multifractal sets, i.e. these equations would describe dynamics of
physical values in the time and the space with fractional dimensions.The
Minkowski-like space-time with fractional (fractal) dimensions for the
case $d_{t}\sim 1$ can be defined on the flat continuous Minkowski
space-time (that is, the measure carrier is the Minkowski space-time
${\mathcal R}^{4}$). These equations can be reduced to the well known
equations of the physical theories for small energy densities, or, which
is the same, for small forces ($d_{t} \to 1$) if we neglect the
corrections arising due to fractality of space and time dimensions (a
number of examples from classical and quantum mechanics and general theory
of relativity were considered in \cite{kob1}).For statistical systems of
many classical particles the GFD help to describe an influence of fractal
structures arising in systems on behavior of distribution functions.

\section{Generalized Newton Equations}

Below we write down the modified Newton equations generated by the
multifractal time field in the presence of gravitational forces
only
\begin{equation}    \label{18}
D_{-,t}^{d_{t}(r,t)}D_{+,t}^{d_{t}(r,t)} {\bf r}(t)=
D_{+,r}^{d_{r}}\Phi_{g}({\bf r}(t))
\end{equation}
\begin{equation}    \label{19}
D_{-,r}^{d_{r}}D_{+,r}^{d_{r}}\Phi_{g}({\bf r}(t))+
\frac{b_{g}^{2}}{2}\Phi_{g}({\bf r}(t))=\kappa
\end{equation}
In (\ref{19}) the constant $b_{g}^{-1}$ is of order of the size of
the Universe and is introduced to extend the class of functions on
which generalized fractional derivatives concept is applicable.
These equations do not hold in closed systems because of the
fractality of spatial dimensions, and therefore we approximate
fractional derivatives as $D_{0+}^{d_{{\bf r}}}\approx {\bf
\nabla}$. The equations complementary to (\ref{18})-(\ref{19})
will be given in the next paragraph. Now we can determine $d_{t}$
for the distances much larger than gravitational radius $r_{0}$
(for the problem of a body's motion in the field of
spherical-symmetric gravitating center) as (see (\ref{9}) and
\cite{kob1_1} for more details)
\begin{equation}    \label{20}
d_{t}\approx 1+\beta_{g}\Phi_{g}
\end{equation}
Neglecting the fractality of spatial dimensions and the
contribution from the term with $b_{g}^{-1}$, and taking
$\beta_{g}=2c^{-2}$), from the energy conservation law
(approximate since our theory and mathematical apparatus apply
only to open systems) we obtain
\begin{eqnarray}   \label{21} \nonumber
& &\left[1-\frac{2\gamma M}{c^{2}r}\right] \left(\frac{\partial
r(t)}{\partial t}\right)^{2} \\ & &+ \left[1-\frac{2\gamma
M}{c^{2}r}\right]r^{2} \left(\frac{\partial \varphi(t)}{\partial
t}\right)^{2}-\frac{2mc^{2}}r=2E
\end{eqnarray}
Here we used the approximate relation between generalized fractional
derivative an usual integer-order derivative (\ref{8}) with $a=0.5$ and
notations corresponding to the conventional description of motion of mass
$m$ near gravitating center $M$. The value $a=0.5$ follows from the
regularization method used and alters if we change the latter.
Eq.(\ref{21}) differs from the corresponding equation in general theory of
relativity by presence of additional term in the first square brackets.
This term describes velocity alteration during gyration and is negligible
while perihelium gyration calculations. If we are to neglect it,
Eq.(\ref{21}) reduces to the corresponding classical limit of equations of
general relativity equation. For large energy densities (e.g.,
gravitational field at $r<r_{0}$) Eqs.(\ref{5}) contain no divergences
\cite{kob1}, since integrodifferential operators of generalized fractional
diferentiation reduce to generalized fractional integrals (see (\ref{1})).

Note, that choosing for fractional dimension $d_{\br}$ in GFD
$D^{d_{\br}}_{0+}$ Lagrangian dependence in the form $L_{\br,i}\approx
L_{t,i}$ gives for (\ref{15}) additional factor of $0.5$ in square
brackets in (\ref{21}) and it can be compensating by fitting factor
$\beta_{g}$.

\section{Fields arising due to the fractality of spatial
dimensions ("temporal" fields)}

If we are to take into account the fractality of spatial
dimensions ($d_{x}\ne 1,\;d_{y}\ne 1,\;d_{z}\ne 1$),
Eqs.(\ref{15})-(\ref{17}), we arrive to a new class of equations
describing certain physical fields (we shall call them "temporal"
fields) generated by the space with fractional dimensions. These
equations are quite similar to the corresponding equations that
appear due to fractality of time dimension and were given earlier.
In Eqs.(\ref{8})-(\ref{10}) we must take $x=\br,/; \alpha=\br$ and
fractal dimensions $d{\br}(t(\br),\br)$ will obey (\ref{12}) with
$t$ being replaced by $\br$. For example, for time $t(\br(t),t))$
and potentials $\Phi_{g}(t(\br),\br)$ and $\Phi_{e}(t(\br),\br)$
(analogues of the gravitational and electric fields) the equations
analogous to Newton's will read (here spatial coordinates play the
role of time)
\begin{equation}    \label{23}
D_{-,\br}^{d_{\br}(\br,t)}D_{+,\br}^{d_{\br}(\br,t)}t(\br)=
D_{+,t}^{d_{t}}\left(\Phi_{g}(t(\br))+
e_{r}m_{r}^{-1}\Phi_{e}(t(\br))\right)
\end{equation}
\begin{equation}        \label{24}
D_{-,t}^{d_{t}}D_{+,t}^{d_{t}}\Phi_{g}(t(\br))+
\frac{b^{2}_{gt}}{2}\Phi_{g}(t(\br))=\kappa_{r}
\end{equation}
\begin{equation}        \label{25}
D_{-,t}^{d_{t}}D_{+,t}^{d_{t}}\Phi_{e}(t(\br))+
\frac{b^{2}_{et}}{2}\Phi_{e}(t(\br))=e_{r}
\end{equation}
These equations should be solved together with the generalized
Newton equations (\ref{18})-(\ref{19}) for $\br(t(\br),\br)$.

With the general algorithm proposed above, it is easy to obtain
generalized equations for any physical theory in terms of GFD. From these
considerations it also follows that for every physical field originating
from the time with fractional dimensions there is the corresponding field
arising due to the fractional dimension of space. These new fields were
referred to as "temporal fields" and obey Eqs.(\ref{12})-(\ref{14}) with
$x=\br,\;\alpha=\br$. Then the question arises, do these equations have
any physical sense or can these new fields be discovered in certain
experiments? I wont to pay attention on the next fact: if $L_{t,i}\approx
L_{\br, i}$ no new fields are generated. This is the case when fractal
dimension of time and space $d_{t,/b r}$ can not be divided on
$d_{t}+d_{\b r}$, the time and space fractal sets can not be divided
too.The FD time and space are common and defined by value given by $L_{i}$
(the latter can be chosen in the form of usual Lagrangians in the known
theories).

\section{Can repulsive gravitational forces exist?}

In general theory of relativity no repulsive gravitational forces
are possible without a change of the Riemann space curvature
(metric tensor changes). But in the frames of multifractal time
and space model, even when we can neglect the fractality of
spatial coordinates, from (\ref{23}) it follows (for
spherically-symmetric mass and electric charge distributions)
\begin{equation}   \label{26}
m_{r}\frac{\partial^{2}t(\br)}{\partial {\br}^{2}}= \frac{\partial}
{\partial t}\sum_{i}\Phi_{i}(t)\approx -\frac{m_{r}k_{r}}
{c^{2}t^{2}}\pm\frac{e^{2}}{ct^{2}}
\end{equation}
with accuracy of the order of $b^{2}$. Here $m_{r}$ is the
analogue of mass in the time space and corresponds to spatial
inertia of object alteration with time changing (it is possible
that $m_{r}$ coincides  with ordinary mass up to a dimensional
factor). Eq.(\ref{26}) describes the change of the time flow
velocity from space point to space point depending on the
"temporal" forces and indicates that in the presence of physical
fields time does not flow uniformly in different regions of space,
i.e the time flow is irregular and heterogeneous (see also Chapter
5 in \cite{kob1}). Note, that introducing equations like
(\ref{26}) in the time space is connected with the following from
our model consequences (see (\ref{15})-(\ref{17})) about
equivalence of time and space and the possibility to describe
properties of time (a real field generating all the other fields
except "temporal") by the methods used to describe the
characteristics of space. Below we will show that taking into
consideration usual gravitational field in the presence of its
"temporal" analogues gives way to the existence of gravitational
repulsion proportional to the third power of velocity. Indeed, the
first term in the right-hand side of (\ref{26}) is the analogue of
gravity in the space of time ("temporal" field). Neglecting
fractional corrections to the dimensions  and taking into account
both usual and "temporal" gravitation, Newton equations have the
form
\begin{equation} \label{27}
m\frac{d^{2}\br}{dt^{2}}={\bf F_{r}}+{\bf F}_{t}=-\frac{\gamma mM}
 {{\bf r}^{2}}+\frac{m_{r}k_{r}}{ct^{2}}\left(\frac{d\br}{dt}\right)^{3}
\end{equation}
The criteria for the velocity, dividing the regions of attraction and
repulsion reads
\begin{equation}   \label{28}
\left(\frac{d\br}{dt}\right)^{3}=\left(\frac{\gamma mM} {{\b
r}^{2}}\right)^{-1}\frac{m_{r}k_{r}}{ct^{2}}
\end{equation}
Here $\br(t)$ must also satisfy Eq.(\ref{18}). Introducing
gravitational radii $r_{0}$ and $t_{0}$ (the latter is the
"temporal" gravitational radius, similar to the conventional
radius $r_{0}$), we can rewrite (\ref{21}) as follows
\begin{equation}
\left|\frac{d\br}{dt}\right|=c\sqrt[3]{c\frac{t^{2}}{r^{2}}\frac{t_{0}}{r_{0}}}
\end{equation}
In the last two expressions $r$ is the distance from a body with mass $m$
to the gravitating center, $t$ is the time difference between the points
where the body and the gravitating center are situated,
$m_{r}=m/c,\;\kappa_{r}=\kappa_{t}/c$. If we admit that $r_{0}$ and
$t_{0}$ are related to each other as $r_{0}=t_{0}/c$, the necessary
condition for the dominance of gravitation repulsion will be $c<rt^{-1}$.
It is not clear whether this criteria is only a formal consequence of the
theory or it has something to do with reality and gravitational repulsion
does exist in nature. What is doubtless, that in the frames of
multifractal theory of time and space it is possible to introduce (though,
may be, only formally) dynamic gravitational forces of repulsion (as well
as repulsive forces of any other nature, including nuclear).

\section{The geometrization of all physical fields and forces}

The multifractal model of time and space allows to consider the
fractional dimensions of time $d_{t}$ and space $d_{\br}$ (or
undivided FD $d_{t{\br}}$ as the source of all physical fields
(see (\ref{9})) (including, in particular, the case when flat (not
fractal) Minkowski space-time ${\mathcal R}^{4}$ is chosen as the
measure carrier). From this point of view, all physical fields are
consequences of fractionality (fractality) of time and space
dimensions. So all the physical fields and forces are exist in
considered model of multifractal geometry of time and space as far
as the multifractal fields of time and space are exists. Within
this point of view, all physical fields are real as far as our
model of real multifractal fields of time and space correctly
predicts and describes the physical reality. But since in this
model all the fields are determined by the value of fractal
dimension of time and space, they appear as geometrical
characteristics of time and space (\ref{10}-\ref{12}). Therefore
there exists a complete geometrization of all physical fields,
based on the idea of time and space with (multi)fractional
dimensions, the hypothesis about minimum of functional of fractal
dimensions and GFD calculus used in this model. The origin of all
physical fields is the result and consequence of the appearing of
the fractional dimensions of time and space. One can say that a
complete geometrization of all the fields that takes place in our
model of fractal time and space is the consequence of the inducing
(and describing by GFD) composed structure of multifractal time
and space as the multifractal sets of multifractal subsets $S_{t}$
and $S_{\b r}$ with global and local FD. The fractionality of
spatial dimensions $d_{\br}$ also leads to a new class of fields
and forces (see (\ref{15})-(\ref{17}) with $\alpha=\br$). For the
special case of integer-valued dimensions ($d_{t}=1,\;d_{\br}=3$)
the multifractal sets of time and space $S_{t}$ and $S_{\br}$
coincide with the measure carrier ${\mathcal R}^{4}$. From
(\ref{12}) it follows then that neither particles nor fields exist
in such a world. Thus the four-dimensional Minkowski space becomes
an ideal physical vacuum (for FD $d_{\alpha}>1$ the exponent of
$R^{n}$ has value $n>4$). On this vacuum, the multifractal sets of
time and space ( $S_{t}$ and $S_{\b r}$) are defined with their
fractional dimensions, and it generates our world with the
physical forces and particles.

Now the following question can be asked: what is the reason for
the dependence in the considered model of fractal theory of time
and space of fractionality of dimensions on Lagrangian densities?
One of the simplest hypothesis seems to assume that the appearing
of fractional parts in the time and space dimensions with
dependence on Lagrangian densities originates from certain
deformations or strains in the spatial and time sets of the
measure carrier caused by the influence of the real time field on
the real space field and vise versa (generating of physical fields
caused by deformations of complex manifolds defined in twistor
space is well known \cite{pen}). Assuming then that multifractal
sets $S_{t}$ and $S_{\br}$ are complex manifolds (complex-valued
dimensions of time and spatial points can be compacted),
deformation, for example, of complex-valued set $S_{t}$ under the
influence of the spatial points set $S_{\br}$ would result in
appearing of spatial energy densities in time dimension, that is
generating of physical fields (see \cite{pen}). Fractional
dimensions of space appearing (under the influence of set $S_{t}$
deformations) yields new class of fields and forces (or can also
not yield). It can be shown also that for small forces (e.g., for
gravity - at distances much larger then gravitational radius)
generalized fractional derivatives (\ref{1})-(\ref{2}) can be
approximated through covariant derivatives in the effective
Riemann space \cite{kob1} and covariant derivatives of the space
of the standard model in elementary particles theory \cite{kob2}
(with the corrections taking into account fields generating and
characterizing the openness of the world in whole
\cite{klim,kob3}). All this allows to speak about natural
insertion of the offered mathematical tools of GFD, at least for
$\varepsilon \ll 1$, in the structure of all modern physical
theories (note here, that the theory of gravitation as the theory
of real fields with a spin $2$ is invented in \cite{log}). Note
also, that  number of problems within the framework of the theory
of multifractal time and space (classical mechanics,
nonrelativistic and relativistic quantum mechanics) were
considered in \cite{kob1,kob3,kob4}.

\section{ Concluions }

In our model we postulate the existence of multifractal space and
time and treat vacuum as ${\mathcal R}^{n}$ space which is the
measure carrier for the sets of multifractal time and space.
Fractionality of time dimension leads then to appearing of
space-time energy densities $L(\br(t),t)$, that is generating of
the known fields and forces, and fractionality of space dimensions
gives new time-space energy densities $L(t(\br),\br)$ and a new
class of "temporal" fields. Note, that the roles of $d_{t}$ and
$d_{\br}$ in distorting accordingly space and time dimensions is
relative and can be interchanged. Apparently, one can consider the
"united" dimension $d_{t,\br}$ - the dimension of undivided onto
time and space multifractal continuum in which time and
coordinates are related to each other by relations like those for
Minkowski space, not using the approximate relation utilized in
this paper $d_t{t,\br}=d_{t}+d_{\br}$. Moreover in some cases it
seems to be even impossible to separate space and time variables,
and then $d_{t}$ and $d_{\br}$ can be chosen to be equal to each
other, i.e., there would be only one fractional dimension
$d_{t}=d_{r_{x,y,z}}=1+\sum\beta_{i}L_{i}(\br(t),t;t(\br),\br)$
describing the whole space-time. In this case one would have to
calculate generalized fractional derivatives from the same
Lagrangians, and new "temporal" fields will not be generated.

The considered model of multifractal time and space offers a new
look (both in mathematical and philosophical senses) onto the
properties of space and time and their description and onto the
nature of all the fields they generate. This gives way to many
interesting results and conclusions, and detailed discussion of
several problems can be found in \cite{kob1}. Here we restrict
ourselves with only brief enumerating of the most important ones.

a) The model does not contradict to the existing physical
theories. Moreover, it reduces to them when the potentials and
fields are small enough, and gives new predictions (free of
divergencies in most cases) for not small fields. Though, the
question about applicability of the proposed relation between
fractal dimension and Lagrangian densities still remains open.

b) We consider time and space to be material fields which are the basis of
our material Universe. In such a Universe there exist absolute frames of
reference, and all the conservation laws are only good approximations
valid for fields and forces of low energy density since the Universe is an
open system, defined on certain measure carrier (the latter probably being
the 4-dimensional Minkowski space). Smallness of fractional corrections to
the value of time dimension in many cases (e.g., on the Earth's surface it
is about $d_{t}-1\sim 10^{-12}$) makes possible to neglect it and use
conventional models of the physics of closed systems.

c) The model allows to consider all fields and forces of the real
world as a result of the geometrization of time and space (may be
more convenient the term "fractalization" of time and space) in
the terms of fractal geometry. It is fractional dimensions of time
and space that generate all fields and forces that exist in the
world. The model introduces a new class of physical fields
("temporal" fields), which originates from the fractionality of
dimensions of space. These fields are analogous to the known
physical fields and forces and can arise or not arise depending on
certain conditions. Thus the presented model of time and space is
the first theory that includes all forces in single theory in the
frames of fractal geometry. Repeat once more: the model allows to
consider all the fields and forces of the world as the result of
geometrization including them in FD of time and space. It is
non-integer dimensions of time and space that produce the all
observable fields. The new class of fields naturally comes into
consideration, originating solely from the fractionality of space
dimensions and with the equations similar to those of the usual
fields. The presented model of space and time is the first theory
that allows to consider all physical fields and forces in terms of
a unique approach.

d) Basing on the multifractal model of time and space, one can
develop a theory of "almost inertial" systems
\cite{kob4,kob6,kob7} which reduces to the special theory of
relativity when we neglect the fractional corrections to the time
dimension. In such "almost inertial" frames of reference motion of
particles with any velocity becomes possible.

e) On the grounds of the considered fractal theory of time and
space very natural but very strong conclusion can be drawn: all
the theory of modern physics is valid only for weak fields and
forces, i.e. in the domain where fractional dimension is almost
integer with fractional corrections being negligibly small.

f) The problem of choosing the proper forms of deformation that
would define appearing of fractional dimensions also remains to be
solved. So far there is no clear  understanding now which type of
fractal dimensions we must use, $d_{t}$ and  $d_{\br}$ or
$d_{t,\br}$. Obviously, solving numerous different problems will
depend on this choice as the result of different points of view on
the nature of multifractal structures of time and space.

The author hopes that new ideas and mathematical tools presented
in this paper will be a good first step on the way of
investigations of fractal characteristics of time and space in our
Universe.

\end{document}